\documentclass[12pt]{article}
\usepackage{epsfig}
\begin{document}
\begin{center}
{\Large {\bf Teleparallel gravity on the lattice.
 \\
}}
\vskip-40mm \rightline{\small ITEP-LAT/2003-29} \vskip 30mm

{
\vspace{1cm}
{M.A.~Zubkov$^a$  }\\
\vspace{.5cm}{\it $^a$ ITEP, B.Cheremushkinskaya 25, Moscow, 117259, Russia }}
\end{center}
\begin{abstract}
We consider quantum gravity model with the squared curvature action. We
construct lattice discretization of the model (both on hypercubic and
simplicial lattices) starting from its teleparallel equivalent. The resulting
lattice models have the actions that are bounded from below while Einstein
equations (without matter) appear in their classical limit.
\end{abstract}

\today \hspace{1ex}
keywords: quantum gravity, teleparallel gravity, lattice discretization

\newpage

Renormalizable asymptotic free theories are, doubtless, well - defined at small
distances.  We cannot claim the same in the case of renormalizable theories
that are not asymptotic free and, of course, in the case of the
nonrenormalizable ones. Therefore, it would be quite natural to perform an
attempt to construct quantum theory of gravity requiring that it must be
asymptotic free.

It was recognized long time ago that ultraviolet divergences in the quantum
gravity with the action that contains  squared curvature term can be absorbed
by an appropriate renormalization of the coupling constants
\cite{renormalizable}. This seems to be a good sign although certain problems
are still present. The model considered  in \cite{renormalizable} and related
publications has the following action (being rotated to Euclidean signature):
\begin{equation}
S = \int \{ \alpha (R_{AB} R_{AB} - \frac{1}{3}R^2) + \beta R^2 - \gamma m_p^2
R + \lambda m_P^4\} |E| d^4 x, \label{S1}
\end{equation}
where $|E| = {\rm det} E^A_{\mu}$, $E^A_{\mu}$ is the inverse vierbein, the
tetrad components of Ricci tensor are denoted by $R_{AB}$, and $R$ is the
scalar curvature. The coupling constants $\alpha, \beta, \gamma$ and $\lambda$
are dimensionless while $m_p$ is a dimensional parameter. Linearized theory
(around flat background) contains graviton together with additional tensor and
scalar excitations. The propagator behaves like $\frac{1}{q^4}$ in ultraviolet
while $\alpha, \beta \ne 0$. Tensor excitation is a ghost, which leads to  loss
of unitarity. A complete, nonperturbative, consideration of the asymptotic
states could possibly become a solution of this problem \cite{unitarity}.
Probably, a traditional approach does not work here just because the
interaction between quantum excitations (and their formation as well) is much
more complicated than it is implied when usual perturbation methods are used.
Nevertheless, the perturbation expansion for the Green functions could still
contain an important information. We expect that due to the renormalizability
of the theory this expansion appears to be self - consistent and could be, in
principle, used as an approximation scheme.

However, the unitarity problem is not the only one that is encountered when we
consider the theory with the action (\ref{S1}). Namely, the requirement that
the action (\ref{S1}) is bounded from below leads to the appearance of a
tachyon. This indicates that flat space is not the real vacuum of the model.
The tachyon would disappear if we construct the perturbation expansion around
the background that minimizes (\ref{S1}) \cite{a_free}. In addition to the
ultraviolet divergences the perturbation expansion may also contain infrared
divergences. In order to separate their consideration from the consideration of
the ultraviolet ones we have to use an additional regularization. This can be
done, for example, if the invariant volume $V$ of the space - time manyfold is
kept constant. Then after the usual regularization (say, the dimensional or
lattice regularization) is removed and all the ultraviolet divergences are
absorbed by the redefinition of the coupling constants, each term of the
perturbation expansion appears to be finite. In the theory with fixed invariant
volume the cosmological constant does not influence the dynamics and the action
is bounded from below if ($\alpha \ge 0$, $\beta > 0$, $\gamma\ne 0$) or
($\alpha \ge 0$, $\beta \ge 0$, $\gamma = 0$). The renormalization group
analysis shows \cite{a_free} that at $\alpha, \beta >0$ there exists a region
of couplings such that the theory is asymptotic free in $\alpha$ and $\beta$
while $\gamma$ can be made constant (up to one - loop approximation). We do not
discuss in this paper the possible divergences that may appear in the limit $V
\rightarrow \infty$. Let us mention, however, that similar divergences do
appear in QED but they are compensated by the ejection of soft photons.
Probably, the same mechanism may work here as well.

Unfortunately the classical Newtonian limit cannot be obtained directly from
the action (\ref{S1}) unless it is not bounded from below. However, if we start
from the pure gravity model with the action (\ref{S1}) (with $\lambda = 0$) and
rotate it back to Minkowski signature, the solutions of Einstein equations
would satisfy the appeared classical equations of motion \footnote{They are not
the only solutions of the equations of motion. However, at $\gamma = \lambda =
0$ Einstein spaces minimize the (Euclidean) action. Therefore it could be
interesting to consider the theory with the action (\ref{S1}) such that at some
scale the {\it renormalized} couplings $\gamma$ and $\lambda$ vanish.}. Then
massive point - like objects could be treated, in principle, as space - time
singularities \cite{mass_singularities} and the Newtonian limit appears as an
asymptotic of black hole solutions. Suppose that the line - like singularity is
embedded into the space - time. Then the Einstein equations in empty space lead
to the Einstein equations in the presence of a particle moving along the
mentioned singularity. Its mass is not fixed by the field equations but it is
proved to be constant along the world trajectory \cite{mass_singularities}.
This indicates that
 matter can be introduced into the quantum gravity
theory with the action (\ref{S1}) in such a way that it reproduces the results
of general relativity at $\alpha > 0,\, \beta > 0, \lambda = 0$.

In this paper we suggest the way to construct lattice theory (both on
hypercubic and simplicial lattices) starting from the action (\ref{S1}) with
$\alpha > 0,\,\beta > 0$.  A traditional approach to the discretization of
gravity is based upon its geometrical interpretation in terms of Riemannian
geometry. Therefore the main concepts are metrics and $SO(3,1)$ (or $SO(4)$)
connection. In several schemes the larger group (say, Poincare or de Sitter
groups) are considered. Vierbein together with the $SO(3,1)$ gauge field appear
 as a part of the correspondent connection, and the metrics is composed of the
 vierbein\cite{Loll,Menotti}. Without  the zero torsion constraint the
 correspondent geometry can be
easily transferred to the lattice. However, it is exactly this constraint that
makes the geometry Riemannian. Its implementation on the lattice is difficult
(although possible \cite{unitarity,Menotti}) and it is quite easy to understand
the reason. The matter is that in Riemannian geometry the space - time indices
are allowed to be mixed with the internal ones. Hypercubic discretization
obviously breaks the $SO(3,1)$ ($SO(4)$) symmetry of space - time. This should
lead, consequently, to breaking of the internal $SO(3,1)$ ($SO(4)$) symmetry.
The theory that is a gauge theory in continuum looses its main symmetry while
being transferred to lattice! Therefore, the $SO(3,1)$ symmetry (like the
chiral one) should be treated on the lattice approximately. Unfortunately, the
correspondent models \cite{unitarity,Menotti} have not yet been investigated
numerically.

An alternative approach to the discretization of Riemannian geometry is given
by the Regge calculus \cite{Regge, Regge_R2, Regge_measure}, where space - time
is approximated by the set of simplices glued together. Each simplex is assumed
to be flat. The geometry is defined by the sizes of the simplices. The
correspondent numerical research has been performed extensively for last $20$
years. The theory with the squared curvature action has been investigated in
this approach (see \cite{Regge_R2}). An essential shortcoming of the Regge
approach is the uncertainty of the choice of the measure over the link lengths
\cite{Regge_measure}.

In the Dynamical Triangulation (DT) variant of Regge calculus \cite{DT} the
sizes of simplices are fixed and the way to glue them together is the dynamical
variable. Unfortunately, except for the case $D = 2$ the simplicial manyfold
cannot reproduce flat space. It is usually implied that flat space appears
approximately. But in this case, say, the local lattice squared curvature
action can not tend to the action (\ref{S1}) at $\gamma = \lambda = 0$ in the
continuum limit: the term linear in $R$ should always appear\cite{DTR2}.

Our approach\footnote{For the other lattice formulations of quantum gravity we
recommend the reader to consult \cite{Loll} and references therein. Some of the
constructions mentioned in \cite{Loll} contain the theory defined by the action
(\ref{S1}) as a limiting case or can be deformed in a way to reproduce it. The
exhaustive numerical investigation was performed only for the Regge and DT
approaches \cite{Regge_R2,DT}, and for de Sitter gravity \cite{Smolin,
dS_numerical}. The model that corresponds to the action (\ref{S1}) was
investigated numerically only within the ranges of Regge approach
\cite{Regge_R2} on extremely small lattices. The extensive investigations were,
however, made in its simplified version ($\alpha = 0$) that is not
renormalizable.} is based upon the teleparallel formulation of general
relativity \cite{teleparallel}. The correspondent geometrical construction is
the so - called Wetzenbock space that appears as a limiting case of a more
general concept - Riemann - Cartan space. The latter is a tangent bundle
equipped with the connection from Poincare algebra. Poincare group consists of
the Lorentz transformations and translations. Translational part of connection
can be identified with the inverse vierbein and defines space - time metrics.
The correspondent part of the curvature becomes the torsion of the Lorentz part
of the connection. The Riemannian geometry appears when the torsion is set to
zero. Weitzenbock geometry is an opposite limit: Lorentz part of Poincare
curvature is set to zero while the torsion remains arbitrary. Teleparallel
gravity is the theory of Weitzenbock geometry, i.e. a translational gauge
theory.

If the space - time manyfold is parallelizable (as it is implied in the present
paper), the zero curvature Lorentz connection can be chosen equal to zero.
Therefore the only dynamical variable is the inverse vierbein, treated as a
translational connection. Usually the action in teleparallel gravity is
expressed through the translational curvature (torsion); the space - time and
internal indices can be contracted by the vierbein. The equivalence between
continuum theories of Riemannian and Weitzenbock geometries can be set up if
everything in Riemannian geometry is expressed through the inverse vierbein and
the latter is identified with the translational connection. For example, there
exists a special choice of classical teleparallel gravity with the action
quadratic in torsion that is identical to usual general relativity.

It is worth mentioning that the concept of torsion has already been transferred
to lattice in the ranges of the Poincare gravity as the dislocation on the
simplicial lattice \cite{Poincare_lat}. Lattice discretization of  teleparallel
gravity may, in principle, be constructed starting from  \cite{Poincare_lat} if
the simplicial manyfold is kept correspondent to flat space. In the resulting
model one would in addition specify definition of the vierbein, the functional
measure and the lattice derivative of torsion. Another representation of
lattice teleparallel gravity was considered in \cite{teleparallel_lat}, where
the basic construction is, in essence, the usual simplicial manyfold used in
Regge calculus. Riemann tensor of the simplicial manyfold can be expressed
through the squared derivatives of the inverse vierbein, that are identified
with the lattice Weitzenbock torsion. As a result in \cite{teleparallel_lat}
the component of lattice torsion relevant for the teleparallel equivalent of
the Regge theory with Einstein action is defined. This approach allows to treat
the simplicial Riemannian manyfold as an approximation of the lattice
Weitzenbock space and to express classical Einstein - Regge equations in terms
of torsion. Unfortunately, the extension of the formalism to higher derivative
gravity has not been made.

In the present paper we suggest an alternative way to discretize teleparallel
gravity that is equivalent to the usual quantum gravity theory with the action
(\ref{S1}). In our approach the Weitzenbock geometry is transferred to lattice
directly, without use of any additional structures (such as the simplicial
Riemannian manyfold of the mentioned above approachs of
\cite{Poincare_lat,teleparallel_lat}). Our main variable is the lattice
connection of Abelian gauge group of translations and everything is expressed
through it.

Before proceed with the description of the lattice construction let us remind
briefly the continuum definition of the model under consideration. The inverse
vierbein is denoted by $E_{\mu}^A$ (everywhere space - time indices are denoted
by Greek letters contrary to the tetrad ones) and is regarded as the
translational connection. In the teleparallel formulation Riemann tensor can be
expressed through the translational curvature (torsion) $T_{ABC} =
E^A_{[\mu,\nu]}E^{\mu}_B E^{\nu}_C$:
\begin{equation}
R_{ABCD} = \gamma_{AB[C,D]} + \gamma_{ABF}\gamma_{F[CD]} +
\gamma_{AFC}\gamma_{FBD} - \gamma_{AFD}\gamma_{FBC},\label{R}
\end{equation}
where
$\gamma_{ABC} = \frac{1}{2}(T_{ABC} + T_{BCA} - T_{CAB})$.
Quantum theory is defined by the functional integral for the partition
function:
\begin{equation}
Z = \int D E exp( - S[E]) \label{Z}
\end{equation}
The measure over $E$ is defined now as the measure over the translational gauge
group.

The theory under consideration is diffeomorfism invariant $(x_{\mu} \rightarrow
\tilde{x}_{\mu}; E^A_{\mu} \rightarrow \frac{\partial \tilde{x}_{\mu}
}{\partial x_{\nu} } E^A_{\nu}$). Substituting Faddeev - Popov unity we
therefore are able to fix partially the gauge, in which $|E| = {\rm const}$.
This freezes the fluctuations of the overall invariant volume. In this paper we
disregard these fluctuations as well as the other global properties of the
space - time manyfold. The correspondent Faddeev - Popov
 determinant is independent of $E$. So, we have
\begin{equation}
Z = \int D E exp( - S[E]) \delta(|E| - {\rm const}),
\end{equation}

As it was mentioned, continuum version of the model was shown to be
renormalizable and asymptotic free for the appropriate choice of
couplings\footnote{The choice of the measure $D E$ is known not to affect this
result}. For these values of couplings the action is bounded from below (if the
fixed invariant volume is kept constant). So we hope that the correspondent
lattice Euclidean functional integral can define a self - consistent theory.


We consider either simplicial or hypercubic lattice. For the further
convenience we refer to simplices (hypercubes) as to elements of the lattice.
Space inside each lattice element is supposed to be flat. Form of the lattice
elements is fixed by the set of vectors ${\bf e}_{\mu}$ connecting the center
of the element with the centers of the sides of its boundary. The expression of
${\bf e}_{\mu}$ through elements of the orthonormal frame ${\bf f}_A$ ($A =
1,2,3,4$) (common for all lattice elements) is the basic variable of the
construction. So we have
\begin{equation}
{\bf e}_{\mu}(x) = \sum_A E^A_{\mu}(x) {\bf f}_A
\end{equation}
We imply that not all of the vectors ${\bf e}_{\mu}$ are independent. Namely,
in hypercubic case
${\bf e}_{\mu} + {\bf e}_{-\mu} = 0$ ($\mu = \pm 4, \pm 3, \pm 2, \pm1$),
where ${\bf e}_{\mu}$ and ${\bf e}_{-\mu}$ connect the center of the hypercube
with the opposite sides of its boundary. In the case of simplicial lattice
($\mu = 1,2,3,4,5$) we have
$\sum_{\mu} {\bf e}_{\mu} = 0$,
Also we imply that all sides of element's boundary are ordered in some way.
Independent variables in both cases are denoted by $E^A_{\mu}$ with $\mu =
1,2,3,4$.

Translational curvature (torsion) is attached to the bones (that are 2 -
dimensional subsimplices in simplicial case and plaquettes in hypercubic case).
We consider the closed path connecting the centers of the lattice elements that
contain the given bone. The piece of the path connecting the centers $x$ and
$y$ of neighbor lattice elements  consists of two vectors ${\bf e}_{M_{xy}}(x)$
and ${\bf e}_{M_{yx}}(y)$ that connect $x$ and $y$ with the center of the side
common for the correspondent lattice elements. Thus integer - valued function
$M_{xy}$ is defined on the pairs of centers of neighbor lattice elements. We
denote ${\bf e}_{xy} = {\bf e}_{M_{xy}}(x) - {\bf e}_{M_{yx}}(y)$ and ${\bf
e}_{yx} = {\bf e}_{M_{yx}}(y) - {\bf e}_{M_{xy}}(x)$. For the closed path $x
\rightarrow y \rightarrow z \rightarrow ... \rightarrow w \rightarrow x$ around
the given bone (consisted of $O$ points) the lattice torsion is
\begin{equation}
T_{xyz...w} = \frac{1}{s}({\bf e}_{xy} + {\bf e}_{yz} + ... + {\bf e}_{wx})
\end{equation}
where $s$ is the area inside the path. If we set the distance between the
centers of neighbor elements equal to unity, then $s = 1$ in the hypercubic
case and $s = O\sqrt{\frac{D+1}{D-1}}$ with $D = 4$ in the simplicial case.
 We also denote
$T_{xyz...w} = T^A_{M_{xy} M_{xw}}(x) {\bf f}_A$,
where summation over $A$ is assumed.

Now let us express basis vectors ${\bf f}$ through $\bf e$. We denote
\begin{equation}
{\bf f}_A = \sum_{\mu} F^{\mu}_A(x) {\bf e}_{\mu}(x)
\end{equation}
Here the sum is over all vectors $\bf e$ (not only over the four independent
ones). For the hypercubic case we denote by $\bf E$ the $4\times 4$ matrix
consisted of $E^A_{\mu}$ with positive $\mu$. Then $F$ can be chosen in the
form
$F^{\mu}_A = \frac{{\rm sign}(\mu)}{2}({\bf E}^{-1})^{|\mu|}_A$.
For the simplicial case the situation is a little more complicated. In order to
obtain the symmetric expression let us denote $\bar{E}_{\mu} = [E^1_{\mu} ...
E^4_{\mu}]^T$ and
${\bf E}_{\mu} = [\bar{E}_1 ... \bar{E}_{\mu - 1} \bar{E}_{\mu+1} ...
\bar{E}_{5}]$.
Therefore $F$ can be taken in the form:
$F^{\mu}_A = \frac{1}{5}\sum_{\nu \ne \mu}({\bf E}_{\nu}^{-1})^{\rho(\nu,
\mu)}_A$,
where $\rho(\nu,\mu) = \mu$ if $\nu > \mu$ and $\rho(\nu,\mu) = \mu - 1$ if
$\nu < \mu$ . Now tetrad components of the torsion can be easily defined as
\begin{equation}
T_{ABC}(x) = \sum_{\mu,\nu} F^{\mu}_B F^{\nu}_C T^A_{\mu \nu}(x)
\end{equation}

We also define tetrad components of the lattice derivative of $T$:
$T_{ABC,D}(x) = \sum_{\mu} F^{\mu}_D T_{ABC,\mu}(x)$,
where
$T_{ABC,\,\mu}(x) = T_{ABC}(y) - T_{ABC}(x)$
for $\mu = M_{xy}$.

Formally applying expression (\ref{R}) to the defined lattice variables we
obtain the definition of lattice Riemann tensor. Then Ricci tensor and the
scalar curvature are defined as usual. Lattice action therefore is constructed
as follows:

\begin{equation}
S = \sum_x \{ \alpha (R_{AB}(x) R_{AB}(x) - \frac{1}{3}R(x)^2) + \beta R(x)^2 -
\gamma m_p^2 R \} \label{SL}
\end{equation}

We omit here the volume factor $|E|$ and the term proportional to $\lambda$
since the lattice analogue of the gauge $|E| = {\rm const}$ is implied. It is
necessary to consider the model in this gauge if we want to consider the theory
in constant invariant volume. Lattice analogue of the continuous gauge
condition $|E| = {\rm const}$ is the condition of constant invariant volume of
each lattice element. If we denote ${\bf E} = {\bf E}_5$ for the simplicial
case, the correspondent condition can be chosen in the form
${\rm det}\, {\bf E} - v = 0$
(for both lattices), where $v$ is the constant that is proportional to the
overall volume of the lattice.

Thus ${\bf E}$ is the dynamical variable, all $E^A_{\mu}$ are expressed through
it, and the partition function of the model has the form
\begin{equation}
Z = \int \Pi_x  D{\bf E}(x) {\rm exp} ( - S[{\bf E}]) ,
\end{equation}
where
\begin{equation}
D {\bf E} = (\Pi_{A,B} d {\bf E}^A_B) \delta({\rm det}{\bf E} - v)
\end{equation}

In order to complete the construction  global properties of the discretized
manyfold $\cal M$ should be set up. First of all we should mention that to
consider an open space - time manyfold is impossible since we have not
specified boundary terms of the action. Therefore instead of performing a
correspondent construction (that should be based on rather serious theoretical
considerations) we restrict ourselves to the case of a closed manyfold. The
natural restriction on its topology is that it should be parallelizable.
Otherwise our variables do not describe Weitzenbock geometry.

It is much more useful to describe a case of complicated topology by the
simplicial manyfold than by a hypercubic lattice. On the other hand, the theory
defined on the latter one is well - suited for the investigations in the case
of simple topology. The hypercubic lattice is supposed to be used mainly for
the investigations in the case of torus $T^4$ that corresponds to the usual
periodic boundary conditions. Of course, the simplicial construction described
in this Letter can be used in this case as well.


The theory defined on the simplicial lattice can, in principle, also be used
for the investigation of  fixed complicated topology case and the topology -
changing processes\footnote{Strictly speaking, our simplicial construction is
applicable to the case of parallelizable manyfolds only. In the case of
complicated topology the requirement that the manyfold is parallelizable is not
acceptable. Therefore in this case we should supplement the construction
described in this Letter with taking into account the fixed zero curvature
$SO(4)$ connection (living on the $3$ - simplices) that allows to joint
different pieces of the given simplicial manyfold (each piece has a fixed
frame). Correspondingly, the derivatives (and their discretizations) must be
substituted by  the covariant ones.}. The description of these cases is,
however, out of the scope of the present paper. In this respect we would like
to mention that in general case the topologies of the $4$ - dimensional
manyfolds cannot be classified. In order to deal with the classified topologies
a strong constraint on $\cal M$ should be imposed. A usually implied constraint
is the existence of a spinor structure \cite{Hawking}. In this case the
signature $\tau$ and Euler characteristics $\chi$ characterize manyfolds up to
a homotopy (if $|\tau| \ne \chi - 2$). In order to investigate topology
changing processes the simplicial manyfold has to be dynamical (like in the DT
models). In addition to the topology - preserving $(p,q)$ moves the topology
changing moves should be defined (see, for example, \cite{topology_DT}). An
alternative way to investigate dynamical topology is to find Matrix model that
is equivalent to the simplicial theory \cite{de_Bakker}.

Finally, we would like to summarize our expectations. First of all, asymptotic
freedom means that at small distances perturbation theory could be applied and
the dynamics in weak coupling is described through the exchange of virtual
gravitons, tensor and scalar particles. Bare particles cannot, however, be
treated as real quantum states because the virtual tensor particle is a ghost.
The complete treatment of the bound states should appear as a result of the
nonperturbative investigation. A possible analytical approach to the
nonperturbative dynamics  is the investigation of the topological excitations.
Nontrivial solutions of equations of motion for the action (\ref{S1}) (at
$\alpha, \beta > 0$) exist for any topology\footnote{They are not necessarily
Einstein spaces. However, an indirect connection between them and the well -
studied gravitational instantons \cite{topological_solutions} may exist.}. They
can, in principle, play the same role as instantons in quantum chromodynamics.
In their numerical investigation the simplicial formulation of the model could
be useful. The main method of investigation of the constructed lattice models
is, of course, the numerical simulation. The main expectation here is that
(similarly to the asymptotic free Yang - Mills theory) the continuum limit may
appear  at $\alpha, \beta \rightarrow \infty$.

The author is grateful to I.G.Avramidi for the useful private communication and
to M.I.Polikarpov for the continuous encouragment. This work was partly
supported by RFBR grant 03-02-16941, by the INTAS grant 00-00111, the CRDF
award RP1-2364-MO-02, by Federal Program of the Russian Ministry of Industry,
Science and Technology No 40.052. 1.1.1112.

\end{document}